\documentclass{optica-article}
\journal{opticajournal} 

\articletype{Research Article}

\usepackage{lineno}
\begin{document}
\title{Wave pulses with unusual asymptotical behavior \\
at infinity}
\author{Peeter Saari\authormark{1,2,*} and Ioannis Besieris\authormark{3} }
\address{\authormark{1}Institute of Physics, University of Tartu, W. Ostwaldi 1, 
50411, Tartu, Estonia\\
\authormark{2}Estonian Academy of Sciences, Kohtu 6, 10130 Tallinn, Estonia\\
\authormark{3}The Bradley Department of Electrical and Computer Engineering, 
Virginia Polytechnic Institute and State University, Blacksburg, Virginia 24060, USA}

\email{\authormark{*}peeter.saari@ut.ee} 
\begin{abstract}The behavior of wave signals in the far zone is not only of theoretical 
interest but also of paramount practical importance in communications and 
other fields of applications of optical, electromagnetic or acoustic waves. 
Long time ago T. T. Wu [J. Appl. Phys. \textbf{57}, 2370 (1985)] introduced  models of "electromagnetic missiles" whose decay 
could be made arbitrarily  slower than the usual inverse distance by an appropriate 
choice of the high frequency portion of the source spectrum. Very recent work by 
Plachenov and Kiselev [Diff. Eqs. \textbf{60}, 1634 (2024).] introduced a finite-energy scalar wave solution, different
from Wu's, decaying slower than inversely proportional with the distance. A physical
explanation for the unusual asymptotic behavior of the latter will be given in this 
article. Furthermore, two additional examples of scalar wave pulses characterized 
by abnormal slow decay in the far zone will be given and their asymptotic behavior
will be discussed. A proof of feasibility of acoustic and electromagnetic fields 
with the abnormal asymptotics will be described.
\end{abstract}

\section{Introduction}
It is a textbook truth that solutions of the wave equation decay 
inversely proportional to distance in the far field or the wave zone 
(see, e.g., \cite{friedlander1964radiation,moses1990acoustic}).  
Long time ago Wu \cite{wu1985EMmissiles} showed that the decay of radiatation
from a finite-size antenna can be made arbitrarily slower thanks to
stretching the Fresnel region for a part of the source spectrum  by a careful 
choice of the high-frequency wing of the spectrum. 
However, subsequent experiments did not substantiate claims 
of unlimited slow decay \cite{shen1988experimental}.
A particular study of quasi-missile behavior was undertaken by Shaarawi \textit{et al.} 
\cite{shaarawi2006localized}.
They examined in detail the asymptotic behavior of a particular scalar
localized wave, referred to as double-exponential pulse (DEX).
In the 1980s, closely related to Wu's 
"electromagnetic missiles", another subject dealing with uncommon  solutions
of the wave equation emerged---the so-called localized or non-diffracting or
space-time wave packets possessing exotic properties, such as self-healing,
superluminality, and invariant propagation without spread or decay, 
theoretically up to infinity. By now this subject has grown into a large 
research field of its own, see monograph \cite{LWsbook2} and reviews 
\cite{ziolkowski1989localized, besieris1998PIERS, SaariLorTrans, KiselevReview, 
AbouraddyReview}. Therefore, the decay in the wave zone as a specific subject
in mathematical physics has remained somewhat in the shadows. 
However, in recent years, several papers on the study of asymptotical 
behavior have appeared, see  
\cite{blagovestchenskii2004behavior,plachenov2023energy,
plachenov2024asymptotics} and references therein.

The very recent paper by Plachenov and Kiselev \cite{plachenov2024asymptotics}
(see also \cite{Kiselev2025ArXiv}) introduced a finite-energy solution, different 
from Wu's model \cite{wu1985EMmissiles}, that decays slower than inversely 
proportional to the distance.

They consider the following simple solution
\begin{equation}
\psi \left( R,t\right) =\frac{1}{\left( ct+ict_{s}\right) ^{2}-R^{2}}\;,
\label{splash}
\end{equation}
to the homogeneous wave equation
\begin{equation}
\frac{1}{\rho }\frac{\partial }{\partial \rho }\left( \rho \frac{\partial
\psi }{\partial \rho }\right) +\frac{\partial ^{2}}{\partial z^{2}}\psi =%
\frac{\partial ^{2}}{\partial (ct)^{2}}\psi \;,
\label{waveeq}
\end{equation} 
where $R=\left( \rho ^{2}+z^{2}\right) ^{1/2}$ is the distance of the field
point from the origin, $c$ is the velocity of light (or sound), and $t_{s}$
is a constant that determines the width of the pulse. Here differently from
 \cite{plachenov2024asymptotics} we have used somewhat changed
designations and wrote the wave equation in cylindrical coordinates for the
axisymmetric case which is satisfied by the solutions introduced in the
following sections.

$\psi\left(  R,t\right)  $ is a special case of a splash pulse 
\cite{ziolkowski1985splash}. It is frequently used to model ultrashort 
pulses and constitutes a spherically propagating pulse, while the spherical
surface, where the pulse peaks at $R\approx ct,$ collapses into the 
origin at negative times $t<0$ and expands for $t>0.$ The real part 
$\operatorname{Re}\psi\left(  R,t\right)  $ represents a bipolar
pulse and the imaginary part $\operatorname{Im}\psi\left(  R,t\right)$ 
a unipolar one when the pulse moves in the far field where $R\gg ct_{s}$, 
and \em vice versa \em near the origin where $R<ct_{s}$.

They consider the limit%
\begin{equation}
L=\lim_{t\rightarrow\infty}\left[  ct\,\psi\left(  R,t\right)  \right]
_{R=ct+\Delta}\ ,\label{Lim}%
\end{equation}
where $\Delta$ is zero or an arbitrary (not large) real number, so that
substitution $R=ct+\Delta$ leads to so to speak riding on the top of the pulse
peak or its proximity as $t\rightarrow\infty$. For a wave function in the form
of Eq.~(\ref{splash}) it turns out that $L=1/\left(  2ict_{s}-2\Delta\right)
$, i.e., it is finite.

The limit of the type of Eq.~(\ref{Lim}) plays an important role in various
fields of mathematical physics and it is known that for finite-energy
"well-behaving" solutions of the wave equation the limit $L$ should be finite
(incl. $L=0$), see, e.g., \cite{friedlander1964radiation, moses1990acoustic, blagovestchenskii2004behavior,plachenov2023energy}.

The authors of \cite{plachenov2024asymptotics} showed that for the
primitive (antiderivative) of $\psi\left( R,t\right)$
\begin{equation} \label{Psii}
\begin{split}
\Psi\left(  R,t\right) =\int_{-\infty}^{t}\psi\left(  R,t^{\prime
}\right)  dt^{\prime}&=\\=\frac{1}{2cR}\ln\frac{ct+ict_{s}-R}{ct+ict_{s}%
+R}, 
\end{split}
\end{equation}
which of course also obeys the wave equation, but in contradistinction to the usual 
behaviour of fields in the wave zone, the quantity in Eq.~\eqref{Lim} is diverging: 
\begin{equation}
L=-\left(  1/2c\right)  \lim_{t\rightarrow\infty}\ln
ct=-\infty. 
\end{equation}
They also showed that $\Psi\left(  R,t\right)  $ has finite total
energy, i.e., may not be an unphysical solution. However, both the scalar field 
energy density and Poynting vector have normal asymptotics in the sense of finite
values of the limit of the type in Eq.~\eqref{Lim}.

The general motivation of our article is to draw attention to the phenomenon of 
abnormally slow decay, which is quite unexpected, at least in the case of physically 
feasible waves. Our aim is to find additional examples of wave
pulses with unusual asymptotics, and to reveal the causes of their abnormally slow 
decay. We pay special attention to determining the finiteness of the energy of the 
found pulses because this property is a crucial one for their physical feasibility 
In the next section we consider one of such examples---the primitive 
(antiderivative) of the wave function of a unidirectional pulse. Section~3 
is devoted to another example---the fractional splash pulse which already
itself, without taking its time integral, decays very slowly in the far zone.

The authors in the short communication \cite{plachenov2024asymptotics} 
neither illustrate graphically the behavior of $\Psi\left(  R,t\right) $ nor 
interpret physically its unusual asymptotics. In Section 4 we try to fill in
this gap. 
In Section~5 we discuss some subtleties of the slow decay of the other pulses 
studied in Sections~2 and 3 and consider the physical
feasibility in  optics, electromagnetics, or acoustics of the pulses 
considered in this study. 

\section{Hemispherical unidirectional pulse}
Now we will consider a cylindrically symmetric pulse which is unidirectional
in the sense that all its plane-wave constituents propagate solely into the
hemisphere with the axial component of the wave vector $k_{z}>0$. 
Such pulses have been intensively studied in recent years
\cite{lekner2018unidir,
KiselevUni, IBB, meiePRA23, saari2024backfl, plachenov2024simple}. 
Surprisingly, a unidirectional pulse may exhibit locally negative values 
of the $z$ component of the Poynting vector, i.e., energy flow in the 
direction opposite to the pulse propagation axis $z$. Such an energy 
backflow effect was studied in detail with respect to a class of pulses a
typical representative of which is given by the following expression \cite{meiePRA23}:
\begin{equation} \label{unorig}
\begin{split}
u\left(  \rho,z,t\right)    & =\frac{1}{g\left(  \rho,t\right)  \left[
i\left(  z-iz_{s}\right)+g\left(\rho,t\right)  \right] }, \\
g\left(  \rho,t\right)    & \equiv\sqrt{\rho^{2}-\left(  ct+ict_{s}\right)
^{2}}.
\end{split}
\end{equation}
Here $z_{s}$ is a positive constant that contributes additionally to
the pulsewidth. It is interesting to note that $u\left(  \rho,z,t\right)$
is a double antiderivative, first with 
respect to time and then with respect to the variable $z$, of another 
unidirectional solution, given in Eq.~(5) of Ref.~\cite{meiePRA23} 
(with replacement $z\rightarrow -z$), which in turn is a compact version 
of Eq.~(3.10) derived in Ref.~\cite{lekner2018unidir} by a unidirectional 
Fourier synthesis.

Similarly to $\psi\left(  R,t\right)$ in Eq.~\eqref{splash}, the real 
and imaginary parts of $u\left(  R,t\right)$ represent bipolar and unipolar 
pulses, but in contrast to the spherically symmetric case, in the given case the
pulses "ride" on a collapsing-expanding tube-like surface with a
hemispherical surface inside.

$u\left(\rho,z,t\right)$ has normal asymptotics:
\begin{align}
&  \lim_{t\rightarrow\infty}\left[  ct\,u\left(  \rho,z,t\right)  \right]
_{z=ct+\Delta}=\frac{i}{ct_{s}+z_{s}+i\Delta},\label{uPatz}\\
&  \lim_{t\rightarrow\infty}\left[  ct\,u\left(  \rho,z,t\right)  \right]
_{z=-(ct+\Delta)}=0,\label{upatnegz}\\
&  \lim_{t\rightarrow\infty}\left[  ct\,u\left(  \rho,z,t\right)  \right]
_{\rho=ct+\Delta}=\frac{i}{2ct_{s}+2i\Delta}.\label{uPatrho}%
\end{align}

But how does its primitive behave? The integral over time
from $-\infty$ to $t$ was taken first by setting $t_{s}=z_{s}=0$ and then 
making replacements  $z\rightarrow z-iz_{s}$ and $t\rightarrow t+it_{s}$.
After some algebra, we found the following expression for the primitive, 
where $g\left(  \rho,t\right)$ is given in Eq.~(\ref{unorig}). 

\begin{align}\label{PrU}
U\left(  \rho,z,t\right)    & =\frac{1}{2h\left(  \rho,z\right)  }\left[
U_{1}\left(  \rho,z,t\right)  +U_{2}\left(  \rho,z,t\right)  \right]
,\\
U_{1}\left(  \rho,z,t\right)    & \equiv\ln\frac{ct+ict_{s}+h\left(
\rho,z\right)  }{ct+ict_{s}-h\left(  \rho,z\right)  }, \nonumber \\ & h\left(
\rho,z\right)  \equiv\sqrt{\rho^{2}+ z^{*2}}, \nonumber \\
U_{2}\left(  \rho,z,t\right)  &  \equiv\ln\frac{\left[  ct^{*}z^{*}  
+g\left(  \rho,t\right)  h\left(
\rho,z\right)  \right]  \left[ z^{*}-h\left(  \rho,z\right)  \right]
}{\left[  ct^{*}z^{*}  -g\left(
\rho,t\right)  h\left(  \rho,z\right)  \right]  \left[ z^{*}+h\left(
\rho,z\right)  \right]  },\nonumber
\end{align}
where $ct^{*}$ stands for  $-ict+ct_{s}$ and $z^{*}$ for $ z-iz_{s}$.
 The behavior of $U\left(  \rho,z,t\right)$ is depicted in Figs.~1 and 2. 
  \begin{figure*}[htbp]
\centering
\includegraphics[width=16 cm]{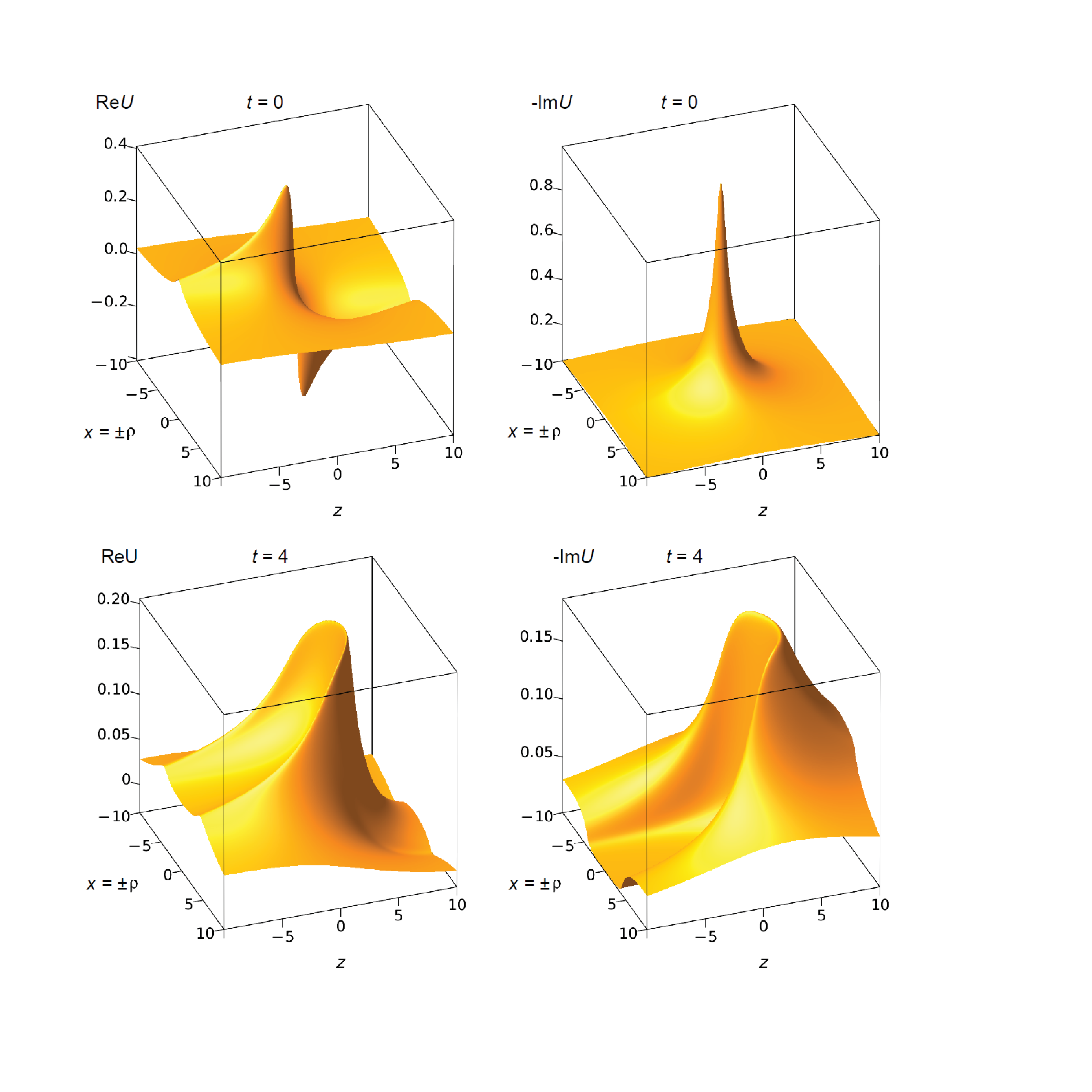}
\caption{The real and imaginary parts
of $U\left(  \rho,z,t\right)$ at the instants $ct=0$ and $ct=4$ ($c\equiv1$).
As the plot is axisymmetric, the axis $x$ represents any axis transverse to the
propagation axis $z$ and, in distinction from the radial coordinate $\rho,$ also takes
negative values. The sign of the imaginary part has been reversed. The pulse width 
parameters are chosen as follows: $ct_{s}=0.3$ and $z_{s}=0.1$. } 
\end{figure*}
\begin{figure*}[htbp]
\centering
\includegraphics[width=16 cm]{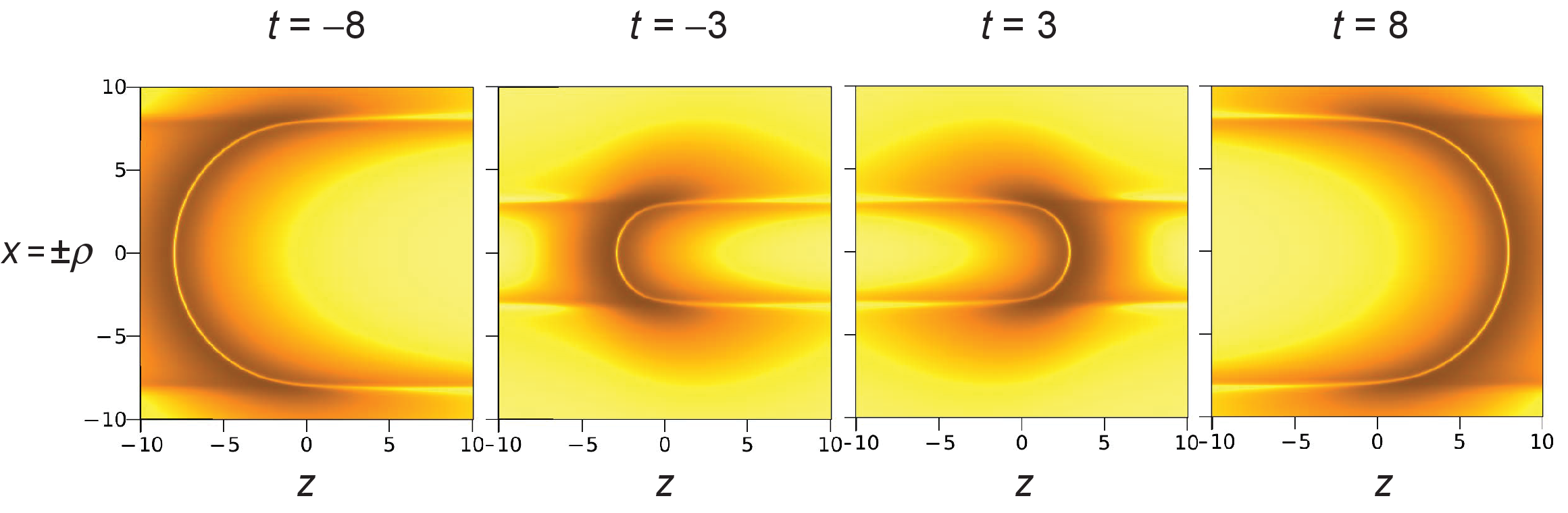}
\caption{"Top view" of the 3D plot of
the real part of $U\left(  \rho,z,t\right)$ at four instants $t$. See caption
of Fig.~1.}
\end{figure*}
  We see that the pulse peaks on the surface of a collapsing-expanding tube with 
 a hemisphere inside it, while the radius of  both is equal
 to $\vert ct \vert$. Hence, $U\left(  \rho,z,t\right)$ behaves similarly to
 $u\left(  \rho,z,t\right)$ from Eq.~(\ref{unorig}), the most remarkable difference
 being that the peak of the real part of the latter is abruptly and symmetrically
 bipolar, except at $t=0$ when it is unipolar contrary to Fig.~1. 

$U\left(  \rho,z,t\right)$ has like $\Psi\left(  R,t\right)$ \em abnormal 
\em asymptotics, except when the limiting process runs in the negative direction of 
the axis $z$:
 \begin{align}
&\lim_{t\rightarrow\infty}\left[ ct\,U\left(\rho,z,t\right)  \right]
_{z=ct+\Delta}=\infty~, \label{Upatz} \\
&\lim_{t\rightarrow\infty}\left[ ct\,U\left(\rho,z,t\right)  \right]
_{z=-(ct+\Delta)}=\ln2~, \label{Upat-z}  \\
&\lim_{t\rightarrow\infty}\left[ ct\,U\left(\rho,z,t\right)  \right]
_{\rho=ct+\Delta}=\infty~, \label{Upatrho}
\end{align} 

Another exception where the limit proves to be finite and equals to $\ln2$
is when time $t$ runs to $-\infty$ but $z$ to $+\infty$. In both these 
exceptional cases the pulse peak is not involved into the process, as can be
seen by looking at Figs. 1 and 2.

We calculated the Poynting vector and energy density of 
$U\left(\rho,z,t\right)$ according to known formulas \cite{Wolf1953scalar} (see also 
Eqs.~(1)-(5) of Ref.~\cite{saari2024backfl}) for scalar fields, as well as 
the total energy. The latter is finite. However, both the Poynting vector
and the energy density have normal asymptotics 
in the sense of finite values of the limit of the type in Eq.~(\ref{Lim}) as they 
have to be since both consist of derivatives of $U\left(  \rho,z,t\right)$
which have normal asymptotics.

Despite  the unidirectionality of  $U\left(\rho,z,t\right)$---integration as 
a linear operation preserves the unidirectional spectrum of 
$u\left(\rho,z,t\right)$---
in certain spatial regions the energy flux turned out to be directed opposite 
to the azis $z$, and at some points the velocity of such an energy backflow 
even reached almost the value $-c$. 
 
We calculated the electric and magnetic fields from $U\left(  \rho,z,t\right)$
using the Hertz vector technique and proved that they have normal asymptotics.
This is understandable, as the technique involves taking derivatives of the scalar field.
 
Finally, we found that the imaginary part
of $U\left(\rho,z,t\right)$ is a "strange"  field
\cite{bessonov1981strange,saari2022strange,plachenov2023pulses} in the sense 
that in certain regions the time integral from $-\infty$ to $+\infty$ is not
zero. This is understandable, because the imaginary part is a unipolar pulse.
 
\section{Fractional splash pulse}
It is known that specific members of a general family of splash pulses are derivable
as spectral superpositions of the most investigated localized propagation-invariant
unipolar pulse---the focus wave mode 
\cite{ziolkowski1989localized,besieris1998PIERS,LWsbook2,ziolkowski1985splash}. 
A specific fractional splash pulse is given by the following expression
\begin{align}
f\left(  \rho,z,t\right)    & =\frac{1}{a_{1}+i\left(  z-ct\right)
}\label{forig}\\
& \times\frac{1}{\left(a_{2}-i\left(  z+ct\right)  +\rho^{2}\left(
a_{1}+i\left(  z-ct\right)  \right)  ^{-1}\right)^{3/4}},\nonumber 
\end{align}
where $a_{1}$ and $a_{2}$ are positive parameters. The general form of the 
splash mode pulse is given by the expression in Eq.~\eqref{forig}, with  the 
power $3/4$ in the denominator replaced by $ \nu+1$, with $\nu>-1$. 
The fractional pulse above corresponds to $\nu=-1/4$. We note that in the case of 
DEX pulses \cite{shaarawi2006localized}, which are presented as differences of two 
splash pulses, the slow decay appears also if $\nu<-1$, but such pulses are not 
square integrable and the total energy of corresponding fields diverges.

The behavior of $f\left(  \rho,z,t\right)$ is depicted in Fig.~3. Both the real 
and imaginary parts of the pulse possess a 
collapsing-expanding spherical structure, but a peak appears only in a narrow cone
of one hemisphere (of positive $z$ if $ct>0$ and of negative $z$ if $ct<0$).
The real part has mirror symmetry with respect to time reversal 
$t\rightarrow-t$, while the imaginary part additionally reverses its sign.
Outside the region of the origin, the pulse tops of the real part and the
imaginary part do not coincide: the top  of the former is shifted farther
from the origin $z=0$ than the point $z=ct$ by a generally small distance 
about $2a_{1}/3$, and the top of the imaginary part is shifted closer by the
distance about $a_{1}/4$. This follows from a detailed inspection of 
Eq.~\eqref{forig}, as well as from plots which will be presented in Section 5.
But the most remarkable distinction from the pulses considered in the
previous sections is that the fractional splash pulse itself, without the need
to take its antiderivative, exhibits abnormal asymptotics as the limits of
the type in Eq.~\eqref{Lim} show. 
The limits along different directions turn out to be as follows. 
 \begin{align}
&\lim_{t\rightarrow\infty}\left[ ct\,f\left(\rho,z,t\right)  \right]
_{z=ct+\Delta}=\infty\cdot \frac{\exp{(i\frac{3\pi}{8})}}{a_{1}+i\Delta}~, \label{flim1} 
\\
&\lim_{t\rightarrow\infty}\left[ ct\,f\left(\rho,z,t\right)  \right]
_{z=-ct-\Delta}=\frac{i}{2(a_{2}+i\Delta)^{3/4}}~, \label{flim2}  \\
&\lim_{t\rightarrow\infty}\left[ ct\,f\left(\rho,z,t\right)  \right]
_{\rho=ct+\Delta}=\frac{i}{(a_{1}+a_{2}+2i\Delta)^{3/4}}~, \label{flim3}\\
&\lim_{t\rightarrow\infty}\left[ ct\,f\left(\rho,z,t\right)  \right]
_{\rho=ct+\Delta,\, z=ct+\Delta}=0~. \label{flim4}
\end{align} 
Eq.~\eqref{flim1} shows that 
the peaks of both the real and imaginary parts decay slower than the 
normal decay $\sim  1/z$ when one moves with the pulse top along the positive 
direction of the axis $z$ towards infinity. How they approach infinity will be 
discussed in Section 5. As was mentioned above, the
tops of the pulses of the real and imaginary parts of $f\left( 
\rho,z,t\right)$ are shifted from the point $z=ct$, 
and therefore the value of $\Delta$ affects differently the limits of the parts. 
If $ct>0$ and $z>0$, a simple choice is $\Delta=0$ in which case the running point
is simultaneously close to the tops of the pulses of both the real and 
imaginary parts and the exponent in Eq.~\eqref{flim1} shows that at this point both
$ct\, \Re f\left( \rho,z=ct,t\right)$ and $ct\, \Im f\left( \rho,z=ct,t\right)$ 
approach infinity in the same way.
\begin{figure*}[htbp]
\centering
\includegraphics[width=15 cm]{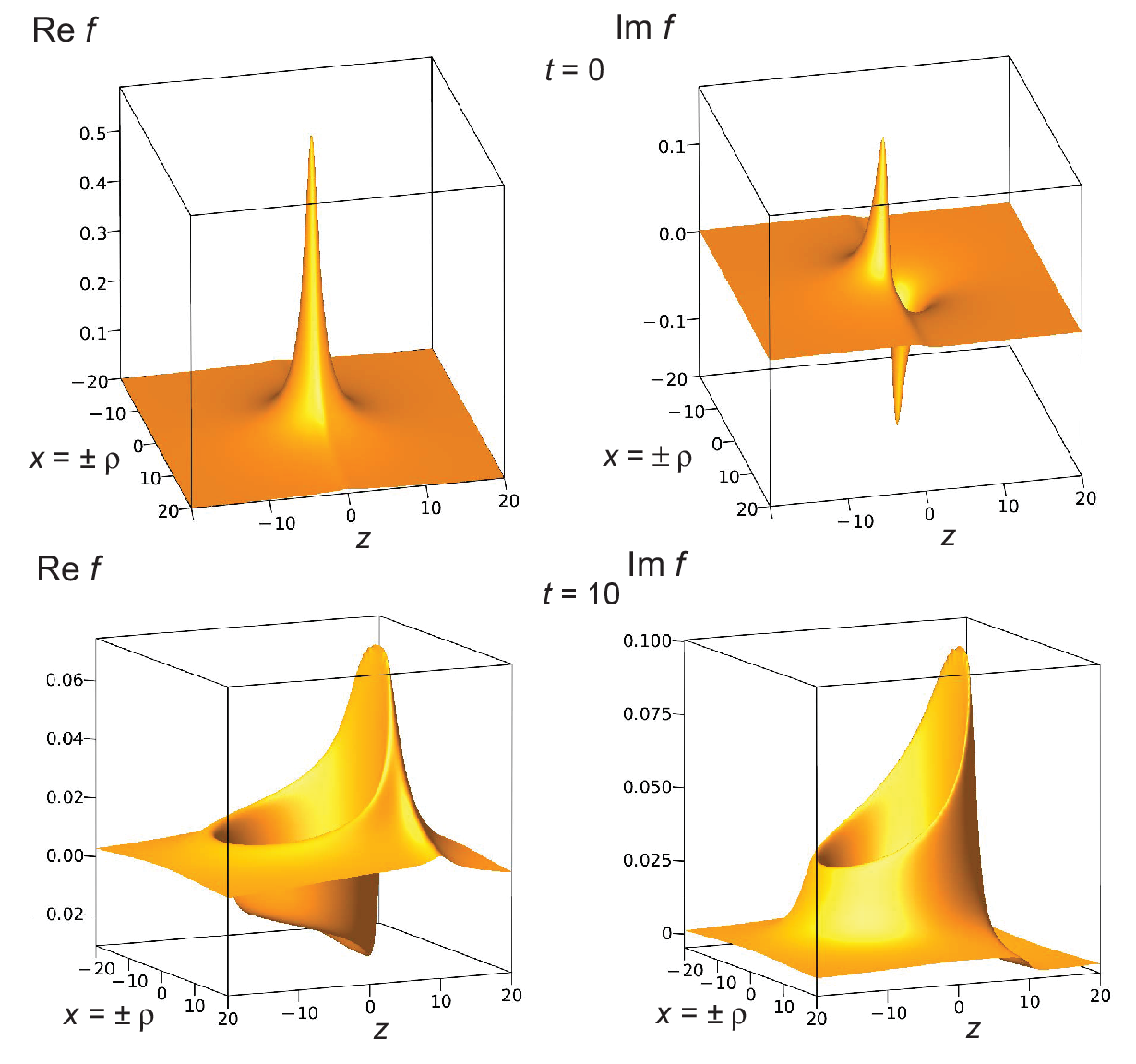}
\caption{The real and imaginary parts
of $f\left(  \rho,z,t\right)$ at  $ct=0$ and $ct=10$. At positive
times, the plot has mirror symmetry $z\rightarrow-z$
The pulse width parameters are $a_{1}=1$ and $a_{2}=2$. See also caption
of Fig.~1.}
\end{figure*}

Eqs.~\eqref{flim2}-\eqref{flim4} state that if $t\rightarrow +\infty$ in 
other direction than along the positive axis $z$, the pulse exhibits
normal asymptotics. Even if one modifies Eq.~\eqref{flim4} by the replacements
$\rho=ct\, \sin\alpha +\Delta,\, z=ct\, \cos\alpha +\Delta$, where $\alpha \approx 0$, 
in order to direct the limit taking almost along the axis $z$, 
the expression of the limit remains finite. This is explained by the relative narrowing 
of the transverse width of the pulse with the propagation distance $z=ct$ and the
remarkable feature that only the top of the pulse decays abnormally with distance. 
Specifically, if we calculate in 3D the solid angle $\Omega=\pi (HWHM)^2/(ct)^2$ 
formed by a circular area with radius equal to transverse  half width at half
maximum (HWHM) of the pulse peak (of the real
part) in the expanding sphere of radius $ct$, then we get the following numerical results. 
When $ct=10$ as in Fig.~3, then $\Omega \approx 2 \,sr$; when $ct=100$, then
$\Omega \approx 0.2\,$; when $ct=1000$, then $\Omega \approx 0.02\, sr$, and
when $ct=10000$, then $\Omega \approx 0.002\, sr$. Hence, although the value of the
HWHM increases with propagation, the \textit{angular} half-width of the peak of 
the pulse  \textit{decreases}  10-fold if the propagation distance increases 10-fold 
and at infinity turns to zero around the axis $z$.

We calculated the Poynting vector and energy density of 
$f\left(\rho,z,t\right)$. Although individually time and space derivatives 
of the fractional splash pulse exhibit abnormal asymptotic behavior,  the energy
density and the Poynting vector when inserted into Eq.~\eqref{flim1} instead of
$f$ do not result in an infinite limit because both quantities are quadratic 
in time and space derivatives. However, for  energy-type quantities, a normal 
decay is inversely proportional to the \textit{square} of the distance. This 
means that in testing the asymptotic behavior,  $ct$ must be replaced by 
$c^2t^2$ in the expression of the limit. In doing so, we arrive at the 
conclusion that the energy density and the Poynting vector also decay 
abnormally slowly. The total energy is finite. $f\left(\rho,z,t\right)$ is 
square integrable (see Appendix), which is also important from a physical point
of view as will be discussed in Section~5. 

We studied also other fractional splash pulses which are given by the same 
expression in Eq.~\eqref{forig} but with other values of the power instead of
$3/4$ in the denominator. If the power is $1/2$, the pulse has the same 
properties as $f\left(\rho,z,t\right)$, but unfortunately its total energy is 
infinite. 
But if the value of the power is between $1/2$ and $1$, i.e., if 
$\frac{1}{2}<\nu+1<1$,  the pulse has slower than normal decay and at the same 
time its wave function is square-integrable and has finite total energy 
(see Appendix).

\section{Causes of uncommon asymptotical behavior at infinity}
It is known that solutions to the homogeneous wave equation can be expressed by 
a convolution of the density of a fictitious Huygens source (coupled with sink)
and the free-space propagator or the Riemann - Schwinger function
\begin{align}
D_{0}(t,R)  & =G_{+}(t,R)-G_{-}(t,R)=\label{RiSw}\\
& =\frac{c}{4\pi R}\left[~\delta\left(R-ct\right)-
\delta\left(R+ct\right)~\right]~,  \nonumber
\end{align}
where $G_{+}$ and $G_{-}$ are the retarded and advanced Green functions,
respectively. By doing this with the delta-point-like source function 
\begin{equation} 
\rho(\textbf{r},t)=\delta(\textbf{r})\hspace{2 pt} \frac{1}
{2c(t+it_{s})}~,\nonumber
\end{equation}
a field like $\psi\left(  R,t\right)$ in Eq.~(\ref{splash}) 
is expressed as a difference of
converging and expanding spherical waves \cite{saari2001evolution}. Specifically,
 $\psi\left(  R,t\right)$  can be decomposed into these two 
waves simply by elementary algebra, viz.:
\begin{align}
\psi\left(  R,t\right) & =\psi_{+}\left(R,t\right)-\psi_{-}\left(R,t\right),
\label{splash2}\\
 \psi_{+}(R,t)& =\frac{1}{2R}\hspace{2 pt}\frac{1}{ct-R+ict_{s}}~,\label{splexp}\\ 
   \psi_{-}(R,t)& =\frac{1}{2R}\hspace{2 pt}\frac{1}
 {ct+R+ict_{s}}~. \label{splcon}
 \end{align}
 Here $ \psi_{-}$ represents the spherical pulse converging at negative 
 times to the origin and $\psi_{+}$ expanding from it at positive times.
\begin{figure}[htbp]
\centering
\includegraphics[width=8 cm]{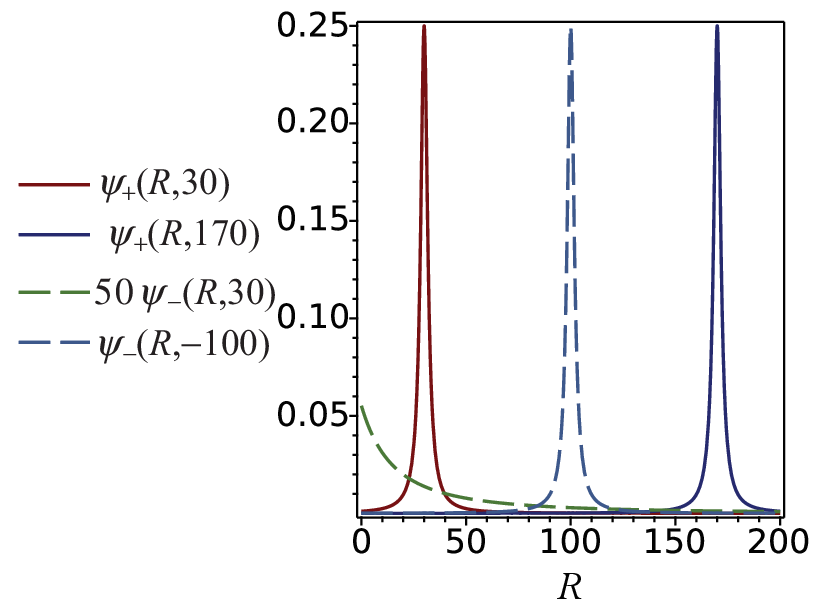}
\caption{Radial dependence of the
imaginary parts of the expanding and converging waves defined 
in Eq.~\eqref{splexp} and Eq.~\eqref{splcon} at instants $ct=30$, 
$ct=170$, and $ct=-100$. All curves have been multiplied by $-R$ and the third 
curve additionally by 50. The pulsewidth parameter $ct_{s}=2$.}
\end{figure}

In Fig.~4 the imaginary parts of the expanding and converging waves 
given by Eq.~\eqref{splexp} and Eq.~\eqref{splcon}, 
respectively, are plotted against the radius $R$ at three
different instants of time. The curves are plotted with reverse sign for
better readability and are scaled---multiplied by $R$---in order
to eliminate the factor $1/R$. Thanks to the scaling we observe that  the 
peaks of the pulses at different time instants are of equal height. This 
correponds to finitness of the limit  in Eq.~\eqref{Lim}, and means that the 
splash pulse has indeed normal asymptotics, as mentioned already in Section 1.
The same holds also for the real parts of the wave functions in Eqs.
~\eqref{splexp} and \eqref{splcon}, except that the pulse of the real part
is bipolar outside of the region of origin at $t=0$, where it is unipolar,
and the pulse of the imaginary part is, on the contrary, bipolar. The 50x 
amplified third curve in Fig.~4 shows the residual tail of the converging wave,
which is very weak at $ct=30$ because its peak is already gone by positive times.  
We shall use Fig.~4 later as a template for graphical illustration for
the explanation of the abnormal asymptotics of the primitive 
in Eq.~\eqref{Psii}.

Quite analogously to Eqs.~\eqref{splash2}-\eqref{splcon}, the primitive of the
splash pulse $\Psi\left( R,t\right)$ can be decomposed into expanding and
converging waves, as can be seen readily from Eq.~\eqref{Psii}
\footnote{If one integrates $\psi_{+}\left(R,t\right)$ and $
\psi_{-}\left(R,t\right)$ over time from $-\infty$ up to any finite value $t$,
then the real parts of both $\Psi_{+}\left(R,t\right)$ and 
$\Psi_{-}\left(R,t\right)$ acquire an additive integration constant
$C=-ln(\infty)$ which cancels out in Eq.~\eqref{Psii2} and therefore does not
affect $\Psi\left(R,t\right)$.}, \textit{viz.}:
\begin{align}
\Psi\left(  R,t\right) & =\Psi_{+}\left(R,t\right)-\Psi_{-}\left(R,t\right),
\label{Psii2}\\
 \Psi_{+}(R,t)& =\frac{1}{2R}\hspace{2 pt}\ln(ct-R+ict_{s})~,\label{Psiiexp}
 \\ 
   \Psi_{-}(R,t)& =\frac{1}{2R}\hspace{2 pt}  \ln(ct+R+ict_{s})~. \label{Psiicon}
 \end{align}
In Fig.~5 the real parts of expanding and converging waves given, 
respectively, by Eq.~\eqref{Psiiexp} and Eq.~\eqref{Psiicon}, as well as of
$\Psi\left( R,t\right)$ from Eq.~\eqref{Psii2}, are plotted against
the radius $R$ at two different instants of time. We no longer deal with the
imaginary parts because their asymptotics are not abnormal. This is 
understandable because, as  integrands in Eq.~\eqref{Psii}, the imaginary
parts of $\psi_{+}\left(R,t\right)$ and $\psi_{-}\left(R,t\right)$ decay with
$t\rightarrow\infty$ as $t^{-2}$, while the real parts as $t^{-1}$.

\begin{figure}[htbp]
\centering
\includegraphics[width=8 cm]{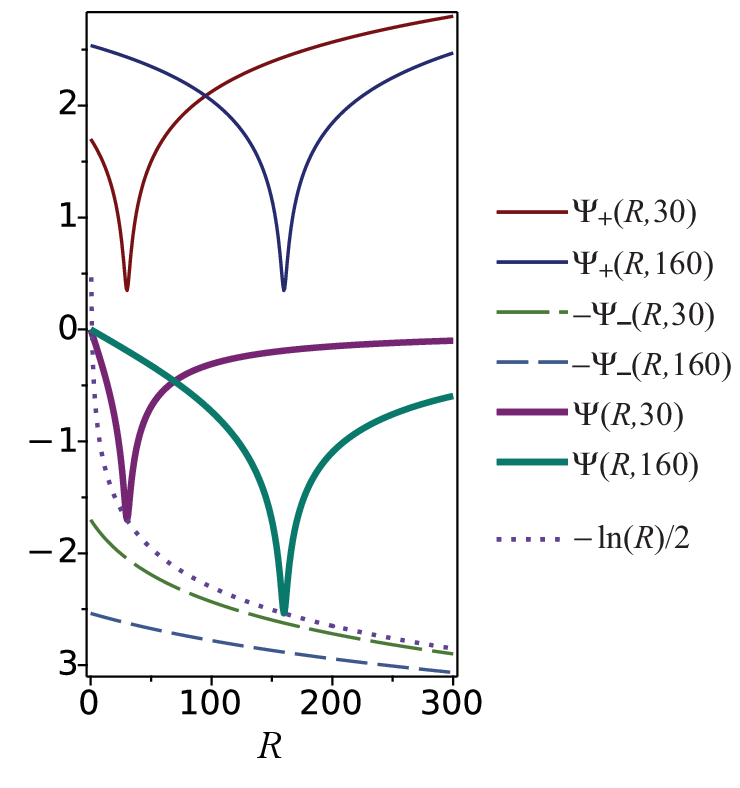}
\caption{Radial dependencies of the
real parts of the waves defined in Eq.s~\eqref{Psiiexp}, \eqref{Psiicon}, 
and \eqref{Psii2} at instants $ct=30$ and $ct=160$. All curves have been
multiplied by $R$. The pulse width parameter is $ct_{s}=2$. For comparison, 
a logarithmically diverging dependence is  shown by the dotted curve,
whereas the factor $1/2$ has been inserted in accordance with the presence of 
the same factor in Eqs.~\eqref{Psiiexp} and \eqref{Psiicon}.}
\end{figure}

The curves are multiplied by $R$ and thus are appropriately scaled for comparison. 
We observe that the peaks of the real part of
$\Psi_{+}(R,t)$ at different time instants are of equal height. This corresponds 
to the finitness of the limit in Eq.\eqref{Lim} and means that the
expanding part of the primitive, taken separately, has normal asymptotics.
But we also observe that if the tails of the converging waves are added---
according to Eq. \eqref{Psii2} with negative sign---the peaks of 
$R\cdot \Re\left[ \Psi\left( R,t\right)\right]$ of the splash pulse grow
logarithmically toward $-\infty$. Indeed, the sum of the ordinate of
the peak at $R=30$ of $R\cdot \Re\left[ \Psi_{+}(R,30)\right]$ and that of
$-R\cdot \Re\left[\Psi_{-}(R,30)\right]$ is exactly equal to the ordinate of
the peak of $R\cdot \Re\left[\Psi(R,30)\right]$ at the same abscissa $R=30$. 
The same can be observed for the curves for the time instant $ct=160$. 

The reason of the abnormal asymptotics of the primitive 
$\Psi(R,t)$ is the presence at $t>0 $ of the tail of the converging 
splash pulse, which according to Eq.~\eqref{Psii} has been integrated from
$ct=-\infty$ up to a given positive time instant. Had we considered a 
non-fictitious radiation source switched on at $ct=0$ and used the 
retarded Green function only, $\Psi_{-}(R,t)$ would be left out 
of Eq.~\eqref{Psii2} and $\Psi(R,t)$ would have normal asymptotics.

Finally, we calculated the energy flux in the pulse $\Psi\left(  R,t\right)$, 
i.e., the Poynting vector and found that for it the limit of type 
Eq.~(\ref{Lim}) is finite, as it has to be since the 
Poynting vector consists of derivatives of $\Psi\left(  R,t\right)$.

\section{Discussion and study of physical feasibility}
The real part of the primitive of the unidirectional pulse has according to 
Eq.~\eqref{Upatz} abnormal asymptotics. With $ct=z\rightarrow \infty$ 
the real part decays as $z^{-1} \ln[z/(ct_{s}+z_{s})]$.
Unfortunately, the expression in Eq.~\eqref{PrU} of the primitive of the 
unidirectional pulse cannot be decomposed into 
expanding and converging parts. However, it is remarkable that in the  
particular case  $z=0$ and $z_{s}=0$  Eq.~\eqref{PrU} 
simplifies substantially, \textit{viz.}:
\begin{equation}\label{PrU00}
U\left(\rho,z,t\right)\mid_{z=z_{s}=0}~ =-\frac{1}{2c\rho} \ln\left(\frac{ct+
ict_{s}-\rho} {ct+ict_{s}+\rho}\right)
\end{equation}    
and coincides with Eq.~\eqref{Psii} if we replace $R\rightarrow\rho$ and
change the sign. Hence, all the results for the primitive of the splash pulse
also apply to the radial evolution of  the primitive of the unidirectional
pulse.

\begin{figure}[htbp]
\centering
\includegraphics[width=7 cm]{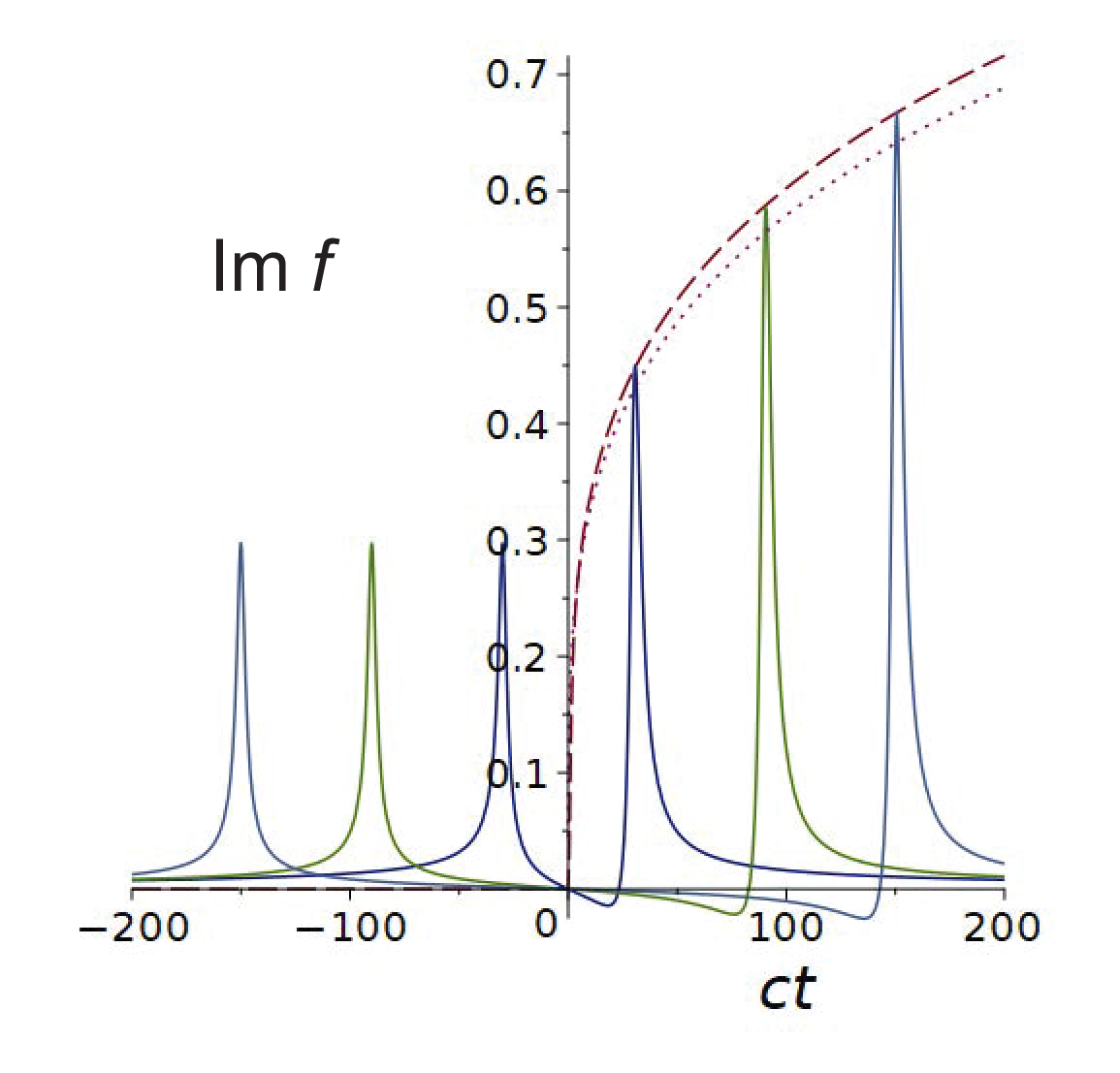}
\caption{Solid curves---time 
dependencies of the imaginary parts (multiplied by $ct$) of the fractional splash 
pulse $f\left(\rho=0,z,t\right)$ at points $z=30$, $z=90$, and $z=150$; 
dashed curve---an analogue of the expression under $lim$ in Eq.~\eqref{Lim},
i.e., $ct\cdot\Im f\left(\rho=0,z=ct+\Delta,t\right)$. For comparison, the 
dotted curve shows a diverging dependence $\sin(3\pi/8)\sqrt[4]{2ct}/2a_{1}
$. The pulse width parameters are $a_{1}=3$ and $a_{2}=2$. $\Delta=-3/4$.
The two last curves are shown only for the region $ct>0$.}. 
\end{figure}

The time dependence of the imaginary part of the fractional splash pulse 
$f\left(\rho,z,t\right) $ at different points on the axis 
$z$ is depicted in Fig.~6. We observe that in the 
region $ct>0$ the strength of the pulses (multiplied by $ct$) grows toward 
infinity, while in the region $ct<0$ it remains constant. The real
part behaves analogously. 

Contrary to the case of $z-$dependence
with fixed time instants, in the case of $ct-$dependence
with fixed locations $z$ the tops  of the curves of the imaginary parts are
shifted to the right (if $t>0$) from the point where $ct$ equals to a fixed
value of $z$.  That is why in order for the dashed curve to touch the tops 
of the peaks, in the expression of $ct\cdot\Im f\left(\rho=0,z=ct+
\Delta,t\right)$  the value of $\Delta$ was set to $\Delta=-a_{1}/4$. If one
sets $\Delta=0$, the dashed curve exactly coincides with the dotted one. 
The latter presents an asymptotical behavior of $ct\cdot f\left(\rho=0,z=ct+
\Delta,t\right)$ according to the following expression
which is directly extracted from the series expansion of the expression in 
Eq.~\eqref{forig}.
\begin{align} 
&ct f\left(\rho=0,z=ct+\Delta,t\right) \propto C(a_{1},\Delta){\sqrt[4]{ct}},
\nonumber\\
C(a_{1},\Delta)&=\frac{1}{(-2i)^{3/4}(a_{1}+i\Delta)},\\
\Re\, C(a_{1},\Delta)&=\frac{\cos(3\pi/8) a_{1}+\sin(3\pi/8)\Delta}{2^{3/4}(a_{1}^2+
\Delta^2)} ,\label{ReC}\\ \Im\, C(a_{1},\Delta)&=\frac{\sin(3\pi/8) a_{1}-
\cos(3\pi/8)\Delta}{2^{3/4}(a_{1}^2+\Delta^2)} \label{ImC}.
\end{align}
The dotted curve has been plotted according to Eq.~\eqref{ImC} as $\Im \,
C(a_{1},\Delta)\sqrt[4]{ct} $ with $\Delta=0$. If one puts here $\Delta=-3/4$,
the dotted curve coincides exactly with the dashed curve.

It follows, then,that the the pulse propagating in the positive direction of the
axis $z$ decays as $\sim 1/(z=ct)^{3/4}$, i.e., slower than the normal decay 
$\sim 1/(z=ct)$. As one can observe in Fig.~6, the pulse acquires 
the abnormal asymptotical behavior already  quite close to the 
origin---after propagating a distance about a ten-fold of its width.

The fractional splash pulse is not unidirectional as its $k_{z}$-spectrum 
shows. This is understandable because it can be presented as a superposition 
of focus wave modes which are bidirectional ('non-causal') pulses. The latter
can be decomposed into forward and backward propagating parts, but the 
corresponding expressions are very  complicated consisting of Lommel 
functions \cite{sheppard2008lommel}. Therefore, we could not make use of the 
superpositional relation between the focus wave mode and the fractional
splash pulse for decomposing the latter.

Not only the total energy of the scalar field $ f\left(\rho,z,t\right)$ is 
\textit{finite}, but also the total energy of an electromagnetic field 
which we proved by numerical double integration and analytical calculations with
the help of spectral representation of fractional splash pulses (see Appendix).

We undertook a thorough proof of the physical feasibility of electromagnetic
fields derivable from the fractional splash pulse $ f\left(\rho,z,t\right)$ with 
the help of a Hertz and a Riemann-Silberstein vector technique. Specifically, we 
chose the Hertz vector as 
$\vec{\Pi}\left(x,y,z,t\right)=\vec{m}f\left(\rho=\sqrt{x^2+y^2},z,t\right)$ where
$\vec{m}$ is a constant directional vector of type $(1,0,0)$ or $(1,1,0)$ or
$(1,i,0)$ \textit{etc}. Excluded was the usual choice of the direction along
the axis $z$ because it turned out to be the only one which did not lead to
abnormally decaying fields. The replacement of $\rho$ was done for more 
convenient  working in Cartesian coordinates. The Riemann-Silberstein vector 
$\vec{F}$ was calculated according to the expression
\begin{equation} \label{R-S}
\vec{F}=\nabla \times \nabla \times \vec{\Pi }+\frac{i}{c}\frac{\partial }{
\partial t}\nabla \times \vec{\Pi }.
\end{equation}
We checked that with $\vec{\Pi}\left(x,y,z,t\right)=\vec{m}f\left(x,y,z,t\right)$ 
the vector $\vec{F}$ indeed obeys the Maxwell equations for free space
\begin{equation} \label{Maxw}
\nabla \times \vec{F}-\frac{i}{c}\frac{\partial }{\partial t}\vec{F}=0,
\quad \nabla \cdot \vec{F}=0.
\end{equation}
Real electric and magnetic fields were calculated by the known relations
\begin{equation} \label{EandB}
\vec{E}=\sqrt{\frac{2}{{{\varepsilon }_{0}}}} \,\Re\vec{F}, \;\;\;
\vec{B}=c^{-1}\sqrt{\frac{2}{{{\varepsilon }_{0}}}} \, \Im\vec{F}.
\end{equation}

The following results were obtained: for all possible choices of $\vec{m}$, 
excluding $(0,0,1)$, the $x$-components and $y$-components of electric and
magnetic fields both have abnormal asymptotics when $z\rightarrow\infty$ while the
$z$-components do not. Hence, $E_x$, $E_y$, $B_x$, and $B_y$ behave 
similarly to $ f\left(\rho,z,t\right)$, see Eq.s~\eqref{flim1}, \eqref{flim2},
\eqref{ReC}, and \eqref{ImC}. In other words, when propagating in the positive 
direction of the axis $z$ they decay as $\sim 1/(z=ct)^{3/4}$, i.e., slower than 
the normal decay $\sim 1/(z=ct)$. If instead of the power $3/4$ another value
between $1/2$ and $1$ is used in
the definition of the splash pulse, the law of the decay changes correspondingly.

\begin{figure}[htbp]
\includegraphics[width=7 cm]{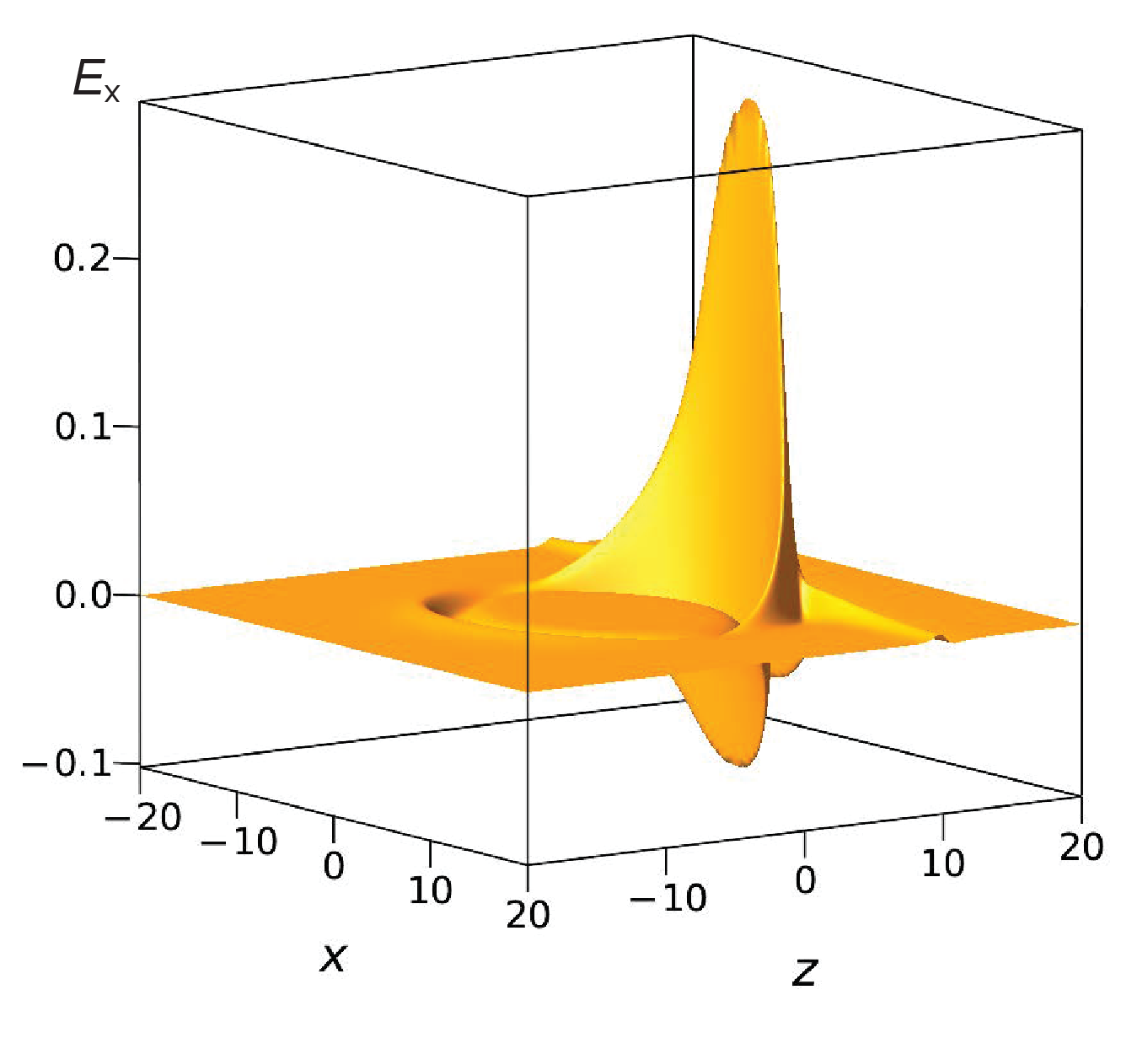}
\caption{The $x$-component of the real
electric field $E_{x}\left(x,y=0,z,t\right)$ at $ct=10$ calculated from the
Hertz vector (1,1,0)$f\left(\rho,z,t\right)$. The pulse width parameters 
are $a_{1}=1$ and $a_{2}=2$. Cf., also,  with Fig.~3.}. 
\end{figure}

Fig.~7 illustrating the behavior of $E_x$ shows that in comparison with 
the scalar-valued wave function in Fig.~3, the pulse's maximum has become more salient,
and the solid angle formed by a circle with radius equal to HWHM of the pulse is 
more than six times smaller than that of the scalar pulse resulting in 
$\Omega \approx 0.0003 \, sr$ if $ct=10000$ (see Section 3).
This is due to the circumstance that according to Eq.~\eqref{R-S} the 
electromagnetic fields are expressed from $f\left(x,y,z,t\right)$ through double
temporal and spatial derivatives which take their highest values in the region of 
the pulse maximum. It is interesting to note that only 
$\partial^2_{z,z}$ and $\partial^2_{t,z}$ lead to terms with abnormal asymptotics. 

In acoustics also the potential $f\left(x,y,z,t\right)$  is not directly observable.
The main physical observables are the  flow velocity $\vec{v}=-\nabla \, \Re 
f\left(x,y,z,t\right)$ and the  excess pressure 
$p=\rho \, \partial_t \,\Re  f\left(x,y,z,t\right)$, where $\rho$ is
the mass density. Both have abnormal asymptotic behavior analogous to that of 
$f\left(x,y,z,t\right)$. However, like in the case of electromagnetic fields, these
quantities must be consistent with the basic equations of fluid dynamics--- the
continuity equation and Euler's equation of motion  \cite{lekner2006sound}:
\begin{align}
 \frac{\partial\rho}{\partial t} +\nabla\cdot (\rho \, \vec{v}) &= 0\,, \label{pidev} \\
 \rho\left[\frac{\partial\vec{v}}{\partial t}+(\vec{v} \cdot \nabla )\vec{v}\,\right]
 &= -\nabla\,\rho \, .\label{motion}
\end{align}
We checked that when  $\vec{v}$ and $p$ are calculated to first order, indeed they 
obey Eq.s~\eqref{pidev} and \eqref{motion}, i.e., they are physical observables.

The fact that the fractional splash pulse is not unidirectional does not imply that
its physical feasibility is impossible. Its backward-propagating components can be
suppressed through appropriate adjustments of the parameters $a_1$ and $a_2$. 
A practical approach to generate a unidirectional finite-aperture
approximation of the pulse is using the Huygens approach \cite{shaarawi2006localized}, 
or  following  a method similar to what was done in the experimental realization of 
the focus wave mode \cite{Reivelt2002fwm1,Reivelt2002fwmexper}. As pointed out in
\cite{shaarawi2006localized}, the resulting finite-energy unidirectional fractional 
splash mode pulse  may not retain the abnormal decay along the z-direction as 
$z\rightarrow\infty$.  There will be, however, an intermediate zone, roughly up to the
Rayleigh distance , where  the pulse will be characterized by slow decay  before it 
assumes the  usual $1/z$  behavior in the far field. If the aperture radius $r_a$  is
equal to the focused radius of the pulse at $z=0$,  then the Raylegh diffraction limit
is given by $Z_R=(\omega_{max} r_a^2)/(2c)$, where $\omega_{max}=4c/a_1$ is the maximum
effective angular frequency in the case of the fractional pulse. It is clear, then, 
that for a small value of the free parameter  $a_1$  and a large aperture the region 
of abnormal (slow) decay can be very large. Generally, the larger the aperture, 
the longer the distance over which abnormally slow decay occurs. In a sense, this 
is analogous to an apertured Bessel beam, where the distance of non-diffracting 
propagation is not infinite but still increases with the size of the aperture. 
We can conclude that wave regions situated far from the propagation axis at $z=0$ 
contribute to the phenomenon of abnormally slow decay. 

As to the primitives in Eqs.~\eqref{Psii} and \eqref{PrU}, the electromagnetic 
and acoustic physical fields derived from them possess normal asymptotics. Therefore,
these primitives are mainly of theoretical interest.

\section{Conclusion}
The solutions to the scalar wave equation with unusual asymptotic behavior 
in the far zone considered in this article are clearly of interest in 
mathematical physics, optics, electromagnetics and acoustics. It is remarkable
that there are several solutions that decay
with propagation distance more slowly than the common inverse proportional law, 
yet still possess finite energy. Especially interesting is the
family of fractional splash pulses. Indeed, it can be easily verified that  
if one changes in Eq.~\eqref{forig} the power $3/4$ with any other value between
$1/2$ and 1, the pulse has a slower than normal decay and, at the same time,
its wave function is square-integrable and has finite total energy.
These characteristics are passed on to physical electromagnetic as well as to 
acoustic fields derivable from the scalar wavefunctions of the fractional 
splash pulses. Hence, the fields with abnormally slow decay in the far zone 
can in principle be implemented as near-cycle pulses of radiofrequency, optical,
or acoustical waves.

The solutions presented here complement the extensive body of work on localized, 
nondiffracting, spatiotemporal, and autofocusing waves, which have already found 
applications in fields such as particle manipulation, micromachining, nonlinear 
spectroscopy, data communication and storage, microscopy, etc., up to medical 
diagnostics and therapy. 
Since, e.g., the fractional splash pulses share some essential properties 
with the listed waves and additionally exhibit exceptionally slow intensity decay 
and angular broadening during propagation, they have promising prospects 
for applications in the same fields.

To summarize, we hope that abnormally decaying pulses as an emerging  subfield in the
study and applications of the localized space-time wave packets, which so far is
represented to our best knowledge only by Refs. 
\cite{wu1985EMmissiles,blagovestchenskii2004behavior, 
shaarawi2006localized,plachenov2023energy,plachenov2024asymptotics} and the present 
paper, will attract growing interest.

\section*{APPENDIX.\\ Proof of square integrability and finiteness of total energy of fractional splash pulses}
Here we present a thorough proof of the square integrability of fractional
splash pulses and finiteness of their total energy. We rely on the approach
developed in Refs.\cite{ziolkowski1989localized, Winful1999EMpulses}, but our
derivation of the key integral relation is simpler. 

A general splash pulse is given by a spectral superposition%
\begin{equation}
f_{G}(\rho,z,t)= \int_{0}^{\infty}G_{FWM}(\rho,z,t)F_{\nu}(k)dk\;,\label{A1}%
\end{equation}
where%
\begin{equation}
G_{FWM}(\rho,z,t)=\frac{\exp\left\{  -k\rho^{2}/\left(  a_{1}+i\tau\right)
+ik\left(  z+ct\right)  \right\}  }{a_{1}+i\tau}\label{A2}%
\end{equation}
is the wave function of the focus wave mode, $\tau=z-ct$, and
$a_{1}$ is a positive constant with the dimension of lenght. $F_{\nu}(k)$ is a
spectrum of the form%
\begin{equation}
F_{\nu}(k)=\frac{1}{\Gamma\left(  \nu+1\right)  }k^{\nu}\exp\left(
-a_{2}k\right)  ;\;a_{2},k>0~,\label{A3}%
\end{equation}
which if $\nu=-1/4$ gives the 3/4 fractional splash pulse in Eq.~\eqref{forig}.

Following the approach of Refs.\cite{ziolkowski1989localized,
Winful1999EMpulses}, the square integrability of the function $f_{G}%
(\rho,z,t)$ is determined by finiteness of the right-hand side of the equality%
\begin{equation}
\int\limits_{0}^{\infty}d\rho\rho\int\limits_{-\infty}^{\infty}dz~\left\vert
f_{G}(\rho,z,t)\right\vert ^{2} =\pi\int\limits_{0}^{\infty}dk~e^{2a_{1}k}%
E_{1}\left(  2a_{1}k\right)  \left\vert F_{\nu}(k)\right\vert ^{2},\label{A4}%
\end{equation}
where $E_{1}\left(  x\right)  $ is the exponential integral function of the
first order. For brevity we have omitted here the factor $2\pi$ which comes
from integration over the azimuthal angle. So our task is to prove Eq.~\eqref{A4} and
then to study whether the right-hand side is finite or diverges depending on
the value of $\nu$. Inserting Eq.s~\eqref{A1}-\eqref{A3} into left-hand side of 
Eq.~\eqref{A4}, the integration over $\rho$ with setting $t=0$ (total energy 
does not depend on time), leaves the integration over $z$ into the form
\begin{equation}
I=\frac{1}{2}\int\limits_{-\infty}^{\infty}dz~\frac{\exp\left[  -i\left(
k_{1}-k_{2}\right)  z\right]  }{a_{1}\left(  k_{1}+k_{2}\right)  +i\left(
k_{1}-k_{2}\right)  z}~.\label{A5}%
\end{equation}
We evaluate this key integral by the following transformations. First for the
sake of brevity we denote $\Delta k\equiv k_{1}-k_{2}$ and $\Lambda\equiv
a_{1}\left(  k_{1}+k_{2}\right)  $. One can verify that $I=0$ unless $\Delta
k=0$. But if we divide the integral into two parts, we obtain%
\begin{align}
I_{1}  & \equiv\frac{1}{2}\int\limits_{-\infty}^{0}dz~\frac{\exp\left[
-i\Delta kz\right]  }{\Lambda+i\Delta kz}=\frac{i}{2}\frac{\operatorname{E}%
_{1}\left(  \Lambda\right)  e^{\Lambda}}{\Delta k},\label{A6}\\
I_{2}  & \equiv\frac{1}{2}\int\limits_{0}^{\infty}dz~\frac{\exp\left[
-i\Delta kz\right]  }{\Lambda+i\Delta kz}=-\frac{i}{2}\frac{\operatorname{E}%
_{1}\left(  \Lambda\right)  e^{\Lambda}}{\Delta k}.\label{A7}%
\end{align}
These equalities can be found by packages of scientific calculations, or proved
by changes of variables of integration. For example, Eq.~\eqref{A7} can be proved by 
the chain of the following changes of variables: $z\rightarrow\Lambda t,$
$t\rightarrow-it,$ $i(1+\Delta kt)\rightarrow y,$ $y\rightarrow it$, which
results in
\begin{equation}
-\frac{i}{2}\frac{e^{\Lambda}}{\Delta k}\int\limits_{1}^{\infty}%
\frac{\exp\left(  -\Lambda t\right)  }{t}=-\frac{i}{2}\frac{\operatorname{Ei}%
_{1}\left(  \Lambda\right)  e^{\Lambda}}{\Delta k},~Q.E.D. \label{A8}
\end{equation}
Here, the definition of the exponential integral function  $E_{1}\left(
x\right)  $ has been used. Eq.~\eqref{A6} is proved in the same way.

Now we apply the relation
\begin{equation}
\lim_{\varepsilon\rightarrow0}\frac{1}{x\mp i\varepsilon}=\pm i\pi
\delta(x)+P\frac{1}{x}\label{A9}%
\end{equation}
known for distributions, to the right-hand sides of Eq.s~\eqref{A6} and \eqref{A7}. 
Common notations are used here: $\delta(x)$ is the Dirac delta and $P$ means the
principal value. Applying Eq.~\eqref{A8} with the opposite signs to 
Eq.s~\eqref{A6} and \eqref{A7} and summing them up, the principal values cancel out. 
Finally, returning to the initial designations we obtain
\begin{equation}
I=I_{1}+I_{2} =\pi\delta(k_{1}-k_{2})\operatorname{E}_{1}\left[  a_{1}\left(
k_{1}+k_{2}\right)  \right]  e^{a_{1}\left(  k_{1}+k_{2}\right)  }~.\label{A10}%
\end{equation}
From insertion Eq.s~\eqref{A1}-\eqref{A3} into left-hand side of 
Eq.~\eqref{A4} there 
remains only the integration over $k_{1}$ and $k_{2}$, one of which can be trivially 
carried out thanks to the Dirac delta in the integrand. Denoting the other spectral
variable by $k$, we obtain, finally, the right-hand side of Eq.~\eqref{A4}. In order 
to study its finiteness, we make use of the upper
bound $\exp(x)\operatorname{E}_{1}(x)<\ln(1+1/x)$ (Eq.~5.1.20 in 
Ref.~\cite{handbook}).
Hence, with the help of Eq.~\eqref{A3}, a splash pulse of index $\nu$ is square
integrable if the integral
\begin{equation}
B_{\nu}(a_{1},a_{2}) \equiv\int\limits_{0}^{\infty}dk~\ln(1+1/2a_{1}k) \frac
{\pi}{\Gamma^{2}\left(  \nu+1\right)  }k^{2\nu}\exp\left(  -2a_{2}k\right)
\label{A11}%
\end{equation}
is finite. With values $-1/2<\nu\leq0$ for the integral in Eq.~\eqref{A11} 
closed-form expressions can be found, which are more or less complicated 
combinations of elementary and special functions that have finite real values if 
$a_{1},a_{2}$ are positive  parameters. 
If $v\leq-1/2,B_{\nu}(a_{1},a_{2})$ diverges, i.e., the 1/2-fractional pulse is 
not square integrable, although it exhibits abnormally slow decay with distance.

Along these lines, also, finiteness of the scalar total energy of the fractional
pulses with $-1/2<\nu<0$ and of electromagnetic and acoustic fields derived from them 
can be established. However, since  these energies contain spatial and temporal
derivatives of the wave function $f_{G}(\rho,z,t),$ they are definitely finite
without the need to calculate them, because derivatives mean multiplication by
$k$ in the spectral domain, and thus remove the singularity at $k=0$ in 
Eq.~\eqref{A11} making the finiteness of it obvious.

Square integrability and finiteness of the total energies, as well as the
independence of them on time, have also been proved by direct numerical 2D integrations.
For example, numerical integration in the left-hand side of Eq.~\eqref{A4}, 
multiplied by $2\pi$ (due to the angular integral), in the case of a particular 
3/4-fractional pulse  ($\nu=-1/4,~a_1=1,~a_2=2$) yields a numerical result 29.041 
independently from time instants $t=0,10,100$. For this case, the Mathematica package 
allows to find a closed-form expression for the right-hand side of Eq.~\eqref{A4}, 
whose numerical value turns out to be equal to 29.041 as well.  For the same case, 
the numerical value of the complicated closed-form expression of the integral in 
Eq.~\eqref{A11}, multiplied also by $2\pi$, turns out to be 33.206, i.e., slightly 
larger as it should be for an upper bound.

For the sake of completeness, we checked the finiteness of the total energy of the scalar
field $f\left(  \rho,z,t\right)$ also analytically. For this the modulus squared in 
the left-hand side of Eq.~\eqref{A4} was replaced by the scalar energy density 
\begin{equation}
   w=\frac{1}{2}\left\vert \frac{\partial}{\partial ct}f\left(  \rho,z,t\right)
\right\vert ^{2}+\frac{1}{2}\left\vert \nabla f\left(  \rho,z,t\right)
\right\vert ^{2}\label{A12}
\end{equation}
and the right-hand side by the same quantity expressed through the spectral 
representation Eqs.~\eqref{A1}-\eqref{A2}. In the case of $\nu=-1/4$ the spectral 
integration in the righ-hand side results in a combination of $\sinh^{-1 }$ and gamma 
functions, the numerical value of which for specified values of parameters 
($a_1=1,~a_2=2$) turns out to be equal to the result of numerical integration 
on the left-hand side, i.e., which proves the finite value of the total energy.

Finiteness of the total energy of the electromagnetic fields, derived from 
$f\left(  \rho,z,t\right)$ by 
the Hertz vector technique, was also proved analytically with the help of Eq.~(B14)
in \cite{Winful1999EMpulses}. Again, this conclusion is natural because 
the EM field expressions contain double derivatives of $f\left(  \rho,z,t\right)$.

\begin{backmatter}

\bmsection{Funding}
No funding was used for the presented research.

\bmsection{Disclosures}
The authors declare no conflicts of interest.

\bmsection{Data availability} No experimental data were generated or analyzed in the presented research.

\end{backmatter}
\bibliographystyle{opticajnl}
\bibliography{Refs4AbnormAsympt.bib}

\end{document}